# Splitting of the universality class of anomalous transport in crowded media

Markus Spanner,[1] Felix Höfling,[2,3] Sebastian C. Kapfer,[1] Klaus R. Mecke,[1] Gerd E. Schröder-Turk,[4] and Thomas Franosch[5,*]

[1]*Institut für Theoretische Physik, Friedrich-Alexander-Universität Erlangen–Nürnberg, Staudtstraße 7, 91058 Erlangen, Germany*
[2]*Fachbereich Mathematik und Informatik, Freie Universität Berlin, Arnimallee 6, 14195 Berlin, Germany*
[3]*Max-Planck-Institut für Intelligente Systeme, Heisenbergstraße 3, 70569 Stuttgart, Germany, and*
*IV. Institut für Theoretische Physik, Universität Stuttgart, Pfaffenwaldring 57, 70569 Stuttgart, Germany*
[4]*Murdoch University, School of Engineering and IT, Mathematics and Statistics, Murdoch, Western Australia 6150, Australia*
[5]*Institut für Theoretische Physik, Leopold-Franzens-Universität Innsbruck, Technikerstraße 21A, A-6020 Innsbruck, Austria*
(Dated: January 21, 2016)

We investigate the emergence of subdiffusive transport by obstruction in continuum models for molecular crowding. While the underlying percolation transition for the accessible space displays universal behavior, the dynamic properties depend in a subtle non-universal way on the transport through narrow channels. At the same time, the different universality classes are robust with respect to introducing correlations in the obstacle matrix as we demonstrate for quenched hard-sphere liquids as underlying structures. Our results confirm that the microscopic dynamics can dominate the relaxational behavior even at long times, in striking contrast to glassy dynamics.

*Introduction.*—The basic paradigm of complex transport in disordered structures was formulated originally by H. A. Lorentz [1] as the motion of a tracer particle in a random medium of independently distributed arrested scatterers. Besides being a testing ground of kinetic theory [2], the Lorentz model has found a fertile soil in many applications, for example electrical conductivity due to impurity scattering [3], hydrogen storage in hierarchically structured porous materials [4], molecular sieving [5, 6], and flow in porous media [7, 8], also in connection to oil recovery [9]. A striking prediction is the emergence of subdiffusive motion, which generically occurs in the crowded world of biological cells [10–14]. Concomitantly, the model exhibits a localization transition [15–17], which has also been found in fluids confined to porous host structures [18–20], and, at intermediate time scales, also in nanoporous silica melts [21–23] or in size-disparate mixtures [24, 25].

At the localization transition, the diffusion coefficient vanishes due to subdiffusive motion of the tracer on arbitrarily long time scales [15]. This dynamic transition is driven by a continuum percolation transition of the underlying geometry [16, 17], sometimes named *Swiss cheese* model. Upon increasing the excluded volume, the spanning cluster in the accessible space is diluted until, at a critical excluded volume fraction, it becomes self-similar with fractal dimension $d_f \approx 2.53$ for space dimension $d = 3$. Hopping transport on such fractals, coined "the ant in the labyrinth" by P. G. de Gennes for critically diluted lattices [26], is equivalent to the conductivity problem of percolating random resistor networks (RRNs) [27]. The mean-square displacement (MSD) of a tracer on the critical infinite cluster becomes itself self-similar, $\delta r_\infty^2(t) \sim t^{2/d_w}$, where $d_w$ is known as the walk dimension. In the vicinity of the transition, universal scaling laws are expected to hold [28–30], concomitant with a power-law growth of the correlation length as controlled by the exponent $\nu \approx 0.88$ ($d = 3$) [28]. Field-theoretical renormalization group arguments explain that the scaling behavior does not depend on the microscopic details and is characterized by universal exponents. While for structural properties, lattice and continuum percolation belong to the same universality class [31], the dynamic universality class describing transport splits similarly to the celebrated models $A$–$J$ for continuous phase transitions [32, 33]. The reason is that for continuum percolation long-range transport depends on the passage through arbitrarily narrow channels. The transition rates $\Gamma$ through the channels become power-law distributed, $\varrho(\Gamma) \sim \Gamma^{-\alpha}$ for small $\Gamma$. A renormalization group analysis shows that the narrow channels either become irrelevant upon coarse-graining or dominate the critical transport entirely [34]. Explicitly, the walk dimension obeys the exponent relation

$$d_w = \max\{d_w^{\text{lat}}, d_f + [\nu(1-\alpha)]^{-1}\}, \qquad (1)$$

where $d_w^{\text{lat}}$ is the universal exponent for Boolean RRNs and for diffusion on lattices [28, 35].

While in two dimensions the universality class of RRNs takes over [36], simulations for the three-dimensional (3D) ballistic Lorentz model [16, 37] confirm the importance of the narrow gaps. In particular, the prediction by Machta and Moore [38] for the exponent $\alpha = 1 - (d-1)^{-1}$ has been corroborated both for the motion on the infinite cluster as well as the motion averaged over all clusters.

In this Letter, we investigate the robustness of the critical dynamics near the localization transition by relaxing some idealizations of the Lorentz model. First, we use quenched hard-sphere fluids as realistic host structures, replacing the uncorrelated overlapping spheres of the original model and thereby changing the statistics of channel widths for the tracer particle. Second, we gradually change the microscopic dynamics from ballistic to Brownian motion, which affects the transit through narrow channels in the system.

*Model.*—We have equilibrated moderately dense hard-sphere configurations consisting of 3 to 5 million particles of diameter $\sigma_{\text{core}}$ at packing fractions $\eta = 0.1$ to $0.4$. Around each of the particle centers, we draw a sphere of diameter

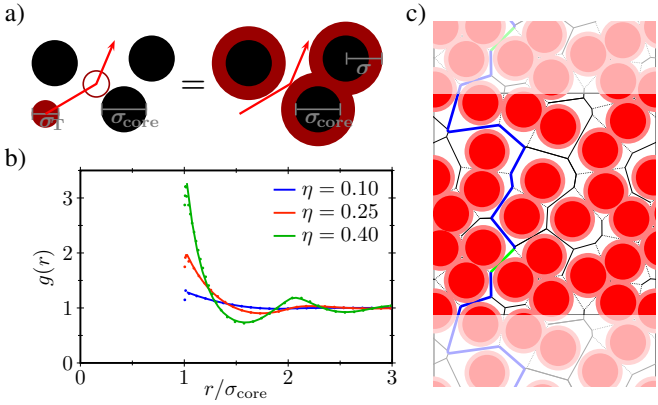

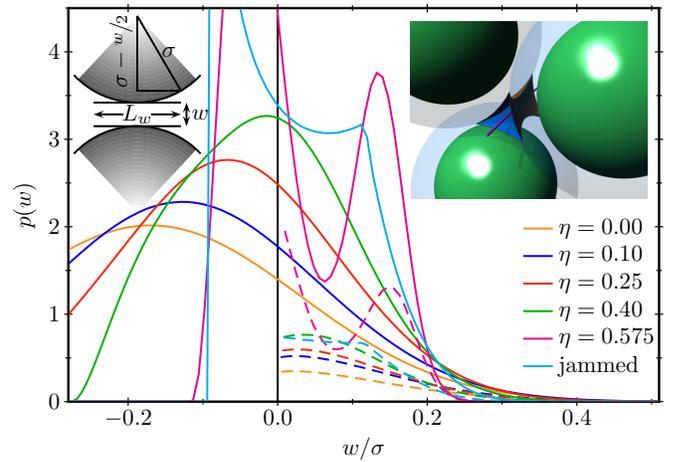

FIG. 1. a) Transport of a tracer particle (diameter $\sigma_T$) in a quenched hard-sphere fluid (left) is mapped to the transport of a point-like tracer in the cherry-pit model (right). Cherry-pit obstacles consist of a hard core of diameter $\sigma_{\text{core}}$ and a shell of width $\sigma - \sigma_{\text{core}}/2 = \sigma_T/2$. Shells of different obstacles can overlap, but the shells can not be penetrated by the point tracer. b) Pair-distribution function $g(r)$ for hard-sphere fluids at packing fractions $\eta = 0.10, 0.25, 0.40$ (solid lines) compared to the Percus–Yevick approximation (dotted lines). c) Voronoi tessellation in the cherry-pit representation ($d = 2$ for illustration). The blue line indicates a path on the percolating cluster containing a narrow channel (green).

FIG. 2. Probability density of channel widths $w$ at the percolation threshold for different packing fractions $\eta$ of the quenched hard-sphere liquid and a jammed hard-sphere configuration. Full lines correspond to all channels, negative widths are inaccessible to the tracer. Dashed lines are the contribution of the spanning cluster. Right inset: Illustration of a channel between 3 neighboring cherry pits. Left inset: Relation between width $w$ and length $L_w \sim \sqrt{w\sigma}$ of a narrow channel, depicted for the 2D case.

$\sigma > \sigma_{\text{core}}$ such that a point-like tracer is confined to the remaining void space—the emerging structure is also known as the cherry-pit model [42]. The overlapping Lorentz model then corresponds to $\sigma_{\text{core}} = 0$, i.e. to vanishing packing fraction $\eta = 0$, see Fig. 1a for an illustration. To ensure proper equilibration of the obstacle matrix, we have monitored the pair-distribution function and compared it to the Percus–Yevick approximation [39] (Fig. 1b). Furthermore we have generated a jammed hard-sphere configuration with resulting packing fraction $\eta \approx 0.6438$ using the Lubachevsky–Stillinger algorithm [40, 41].

Long-range transport in such quenched structures occurs only if the void space percolates through the entire system. In the following, we focus on the critical interaction distance $\sigma$ that marks the percolation transition. The threshold can be determined efficiently by a Voronoi tessellation of the obstacle centers [31]; edges of the tessellation closer than $\sigma$ to an obstacle center are removed. Upon increasing $\sigma$, more and more edges are removed up to the point where the residual network barely spans the simulation box (Fig. 1c). The percolation transition as a function of the packing fraction is discussed in the Supplementary Material [43].

The void space of the cherry-pit structures can be viewed as a network of pores connected by *channels*, where each channel corresponds to an edge of the diluted Voronoi tessellation. For each channel, we define the width $w$ as the distance $\sigma + w$ of its Voronoi edge to the closest obstacle center. Then $w$ corresponds to the radius of the largest sphere that fits through this particular channel—only channels with $w > 0$ can be passed by the tracer. At the percolation transition, the majority of channels is blocked and even the most likely channel width is inaccessible to the tracer (Fig. 2). The distribution is single-peaked for quenched hard-sphere liquids, and structural correlations become more pronounced upon increasing the packing fraction. Then the probability for narrow channels, $0 \leq w \ll \sigma$, increases by a factor of $\approx 2$; simultaneously the distributions become broader and more uniform. The channels associated with the infinite cluster display similar variations. The jammed structure peaks at a minimal channel width corresponding to three particles at contact. Yet, at vanishing channel width, $w \simeq 0$, the distribution shows again a regular variation.

*Probing structural correlations.*—We have monitored the motion of a single tracer particle exploring the void space by event-driven molecular dynamics simulations (see Supplementary Material [43]). Our first set of results is for "ballistic" tracers undergoing specular reflections from the obstacles. The interaction distance $\sigma$ has always been tuned to the percolation threshold.

The MSDs for different packing fractions including the overlapping Lorentz model ($\eta = 0$) and jammed configurations are displayed in Fig. 3. The short-time ballistic motion crosses over smoothly to subdiffusive motion, $\delta r_\infty^2(t) \sim t^{2/d_w}$, extending over more than 7 temporal decades. As a most sensitive test for the anomalous exponent, we have evaluated the local exponents $\gamma_\infty(t) := \mathrm{d}\log \delta r_\infty^2(t)/\mathrm{d}\log t$. Then the data extrapolate nicely (Fig. 3) to the long-time limit $\gamma_\infty(t \to \infty) = 2/d_w$ with the same exponent $d_w = 4.81$ for all densities, which is the well-established value for the overlapping case [16, 17, 37].

*Probing the micro-dynamics.*—Second, we have varied the micro-dynamics of the tracer, starting with deterministic ballistic motion and specular reflections, and then gradually adding randomness. Thereby the narrow channels are probed differ-ently, potentially modifying their respective "conductances". Since structural correlations do not affect the value of $d_w$ as shown above, the simulations are performed only for the overlapping Lorentz model ($\eta = 0$). One possibility to introduce a source of randomness is by changing the scattering rule off the obstacles: In *diffuse scattering* the outgoing direction is random in the accessible hemisphere keeping the magnitude of the velocity fixed, which models a roughness of the scatterer surfaces. Figure 4 reveals that this does not affect the long-time behavior within our time window.

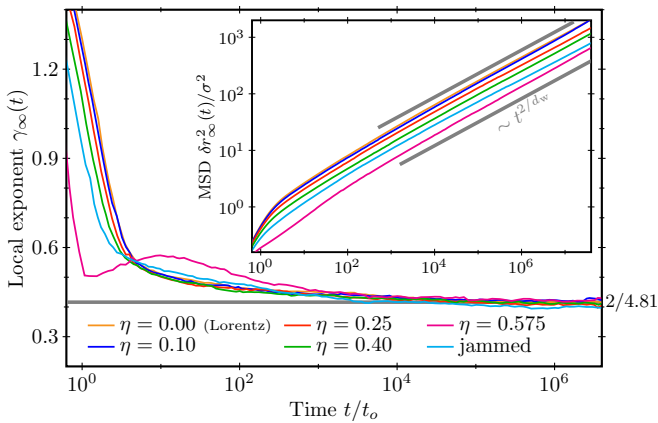

FIG. 3. Local exponents $\gamma_\infty(t) = d \log \delta r_\infty^2(t)/d \log t$ of the ballistic tracer particle at the threshold for different packing fractions $\eta$ (including the Lorentz model $\eta = 0$) of the quenched hard-sphere liquid, and the jammed structures. For the definition of $t_o$, see Ref. [43]. Inset: $\delta r_\infty^2(t)$ for the same systems.

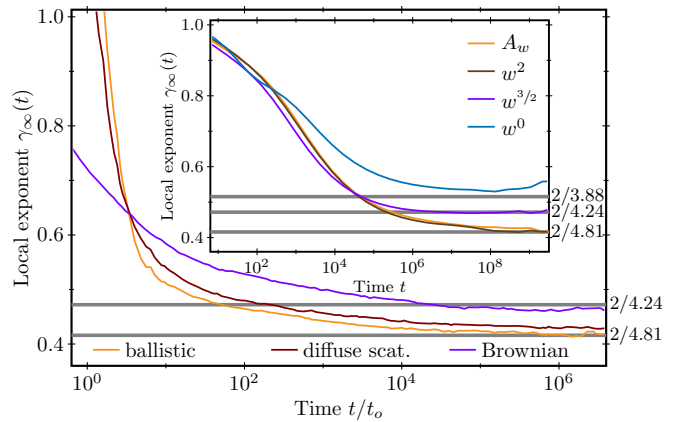

FIG. 4. Local exponents $\gamma_\infty(t)$ in the Lorentz model ($\eta = 0$) at the critical obstacle density for different tracer dynamics: ballistic motion with specular and diffuse scattering; pseudo-Brownian motion with interupt times $\tau_B = 0.005 t_o$. Inset: Local exponents $\gamma_\infty(t)$ for the hopping motion on the Voronoi network using different distributions of random hopping rates.

Next, we consider pseudo-Brownian motion within a coarse-grained scheme where the ballistic trajectory is interrupted after fixed time intervals $\tau_B$ and a random velocity is assigned according to a Maxwell distribution [51]. Then Fig. 4 (see also Supplementary Material [43]) demonstrates that for pseudo-Brownian motion the MSD grows faster and the walk dimension changes to $d_w = 4.24$.

The crossover from the ballistic universality class towards the Brownian one is sensitive to the algorithmic time scale $\tau_B$. A proper mimicry of Brownian motion requires $\tau_B$ to be chosen as small as possible, yet the integration step of the simulation becomes very small and hence computationally unfeasible. Relying on a larger $\tau_B$ is delusive since then the narrowest channels are probed ballistically rather than diffusively. In fact, increasing $\tau_B$ by factors of 10 and 100 yields an apparent convergence of the local exponents to values in between (data not shown).

*Discussion.*—To rationalize our findings we reconsider the arguments [52, 53] for the power-law tail $\varrho(\Gamma) \sim \Gamma^{-\alpha}$ in the distribution $\varrho(\Gamma)$ of small transition rates $\Gamma$. The probability density is connected via $\varrho(\Gamma) d\Gamma = P(w) dw$ to the probability density of a narrow channel. Figure 2, however, shows that $P(w)$ is not singular for narrow channels and attains a finite value $P(w \to 0)$. This insight explains the robustness of the walk dimension with respect to introducing spatial correlations in the host structure.

Therefore a change in the dynamic universality class must originate from the relation between the transition rate and the width of a channel. With each channel of width $w$ and transition rate $\Gamma$ we associate a "conductance" $g(w) \simeq \Gamma/A_w$, respectively a "conductivity" $\sigma(w) \simeq (\Gamma/A_w)L_w$, where $A_w$ is a typical cross-sectional area of a channel and $L_w$ its effective length. For narrow channels $A_w \sim w^{d-1}$, and narrow channels remain narrow on channel lengths $L_w$. From the inset of Fig. 2 one infers $(L_w/2)^2 + (\sigma - w/2)^2 = \sigma^2$, which yields $L_w \sim \sqrt{w\sigma}$ for small channel width $w$.

For ballistic motion, a particle hitting a channel at the correct angle will pass through the channel like a pinball, irrespective of the length of the channel. Then one anticipates that the conductance becomes constant, $g(w \to 0) = const$, for small channel width. This implies that the transition rate becomes proportional to the cross-sectional area, $\Gamma \sim g(w \to 0)w^{d-1}$. Then the rate distribution fulfills $\varrho(\Gamma) \simeq P(w \to 0) dw/d\Gamma \sim \Gamma^{-(d-2)/(d-1)}$ for $\Gamma \to 0$, which yields $1 - \alpha = (d-1)^{-1}$ consistent with the prediction of Machta and Moore [52] and confirmed here and earlier [16, 37].

In contrast, for Brownian particles it is the conductivity that is expected to become constant for small channel widths, $\sigma(w \to 0) = const$. Then the transition rates follow $\Gamma \sim \sigma(w \to 0) w^{d-1} L_w^{-1}$ which yields $\varrho(\Gamma) \simeq P(w \to 0) dw/d\Gamma \sim \Gamma^{-\alpha}$ with $\alpha = (d - 5/2)/(d - 3/2)$. The latter value was suggested for the electrical conductivity in Swiss cheese models [53], yet no connection to Brownian motion was made. Plugging in numbers in Eq. (1) yields $d_w = 4.24$ as found in our simulations.

To corroborate the scenario of how narrow channels are probed we have extracted the Voronoi network of the Lorentz model ($\eta = 0$) and simulated hopping transport on it. The hopping rates between the cells can be measured in principle, and the typical rates as function of the channel width $w$ display the anticipated asymptotics (Fig. S5 in Ref. [43]). Here we follow a different approach and randomly assign hopping rates between nodes connected by unblocked edges according to the following distributions: proportional to, first, the measured cross-sectional area $A_w$ and, second, the anticipated asymptotic behavior of the cross section $\propto w^2$. Third, we have employed hopping rates $\propto w^{3/2}$ in accordance with the expected Brownian probing, and last also hopping rates independent of the channel width as suggested from a lattice RRN. From the inset of Fig. 4, one infers that indeed the dynamic exponent changes with the microdynamics, and extrapolates to the universality class of ballistic and Brownian transport, respectively, or to the conventional lattice RRN, $d_\text{w}^\text{lat} = 3.88(3)$.

*Conclusion.*—We have presented high-precision simulation data for transport in Lorentz models including structural correlations and different micro-dynamics of the tracer. The exponents associated with the anomalous transport are robust with respect to correlations in the arrested host structure. Changing the dynamic rules from ballistic motion to Brownian motion yields a different universality class.

The sensitivity of the asymptotic dynamics on microscopic details is in remarkable contradiction to the dynamic behavior of a glass-forming system close to structural arrest. There one of the hallmarks is that molecular glass-forming liquids with Newtonian dynamics behave identically to colloidal suspensions with Brownian or overdamped dynamics [54–56]. Intuition suggests that the slow dynamics occurs in a rough free energy landscape provided by the configuration rather than phase space, such that ballistic and Brownian dynamics should behave identically. This picture is justified by the escape of a particle from its cage being a local phenomenon even after coarse-graining. In contrast, the slow dynamics on the emergent fractal structure close to the percolation transition is driven by a divergent length scale and the renormalization flow amplifies the role of narrow channels upon coarse-graining.

Evidence for the robustness of the universality class with respect to structural correlations has been collected earlier [57], e.g. by introducing polydisperse obstacles [58]. However, there the simulation was for 2D systems, where the exponent relation Eq. (1) yields $d_\text{w} = d_\text{w}^\text{lat}$ anyway. Similarly, a recent experimental realization [8] of the cherry-pit Lorentz model in $d = 2$ combined with a simulation for soft spheres showed compatibility with the lattice value for $d_\text{w}$. In $d = 3$, structural correlations were investigated before [59] with the same conclusion that the exponent $d_\text{w}$ is robust, yet for dissipative particle dynamics (DPD) and with significantly lower statistical accuracy.

Our data for Brownian motion are the first confirmation of the splitting of the dynamic universality class in Lorentz models. Note that the standard algorithm to generate Brownian motion [51] is specious since the scale-free motion is replaced by a different micro-dynamics at small scales. Yet, narrow channels dominate transport at the percolation threshold and, correspondingly, the narrowest channels are always probed incorrectly. This explains why it has been inferred earlier [17] that (pseudo-)Brownian motion would yield critical exponents identical to the ballistic case. Presumably the same discrepancy is present in Ref. [59].

That the underlying dynamics may be relevant for anomalous transport is anticipated from the wider perspective [10–14]. For example, in the Lévy–Lorentz gas [60, 61] a tracer collides with a diluted fractal structure, resulting in anomalous superdiffusion for ballistic dynamics, while Brownian tracers display ordinary diffusion. Second, the single-file diffusion along 1D channels [62–64] becomes subdiffusive for Brownian dynamics, whereas normal diffusion results for ballistic particle motion. The Lorentz model differs from these examples since the underlying percolation transition constitutes a static critical phenomenon with universal exponents. Different values of the walk dimension $d_\text{w}$ for different microscopic dynamics imply a paradigm-breaking splitting of the universality class.

It is interesting to ask if even more dynamic universality classes exist. For example, if the colloidal realization of the Lorentz model by Skinner *et al.* [8] could be done in $d = 3$, one expects that hydrodynamic effects due to thin lubrication layers dominate transport through the narrowest channels. Similarly, in the complementary molecular dynamics simulations [46], the percolation threshold is set by the energy of the tracer such that the narrow channels are the ones where the energy barely suffices to pass the barriers. Then, by the bottleneck effect, the time to pass these barriers becomes long, giving rise to a singular energy dependence of the conductivity.

This work has been supported by the Deutsche Forschungsgemeinschaft DFG via the Research Unit FOR1394 "Nonlinear Response to Probe Vitrification".

# Supplement to "Splitting of the universality class of anomalous transport in crowded media"

Markus Spanner, Felix Höfling, Sebastian C. Kapfer, Klaus R. Mecke, Gerd E. Schröder-Turk, and Thomas Franosch
(Dated: January 20, 2016)

## A. PERCOLATION THRESHOLD

We have equilibrated hard-sphere configurations consisting of up to $N = 5 \cdot 10^6$ particles of diameter $\sigma_\text{core}$ in a cubic box of linear size $L$ at various packing fractions $\eta = (\pi/6)(N/L^3)\sigma_\text{core}^3$. For a given packing fraction $\eta$, tracer particles can percolate through the system, provided the interaction distance is smaller than a critical value $\sigma = \sigma(\eta)$. We have determined the thresholds relying on a Voronoi tessellation of the obstacle centers [1] using the free voro++ library [2]. The critical interaction distance is converted to a critical dimensionless obstacle density $n_c^* = (N/L^3)\sigma^3$. In particular, the case of the overlapping Lorentz gas $n_c^* \approx 0.838$ [3, 4] is recovered for uncorrelated obstacles, i.e. vanishing hard-sphere diameter: $\sigma_\text{core} = 0$ or $\eta = 0$.

The packing fractions cover the entire fluid phase, the fluid-crystalline coexistence, as well as the fcc crystalline phase up to close packing $\eta = \pi\sqrt{2}/6 \approx 0.74$, see Fig. S1. The small systems, containing $N = 16\,384$ particles, can be equilibrated at a large range of $\eta$ in reasonable time, including packing fractions above the onset of crystallization, yet are sufficiently large to obtain the percolation thresholds with a sample variance of $\lesssim 0.2\%$. The large system size discussed in the main text is for box sizes $L \approx 200\sigma$ in order to minimize finite-size effects for the transport properties at the percolation threshold. Here we checked that the typical excursions of the tracer particle in the time windows investigated are significantly smaller than the system size. Additionally, we also provide systems of an intermediate size for consistency checks.

With increasing packing fraction $\eta$, the increasing order of the particles leads to decreasing overlap, therefore a smaller number of particles is needed to cover an amount of void space that is sufficient to reach the percolation transition. We obtain a good representation for the data in the entire fluid phase by the empirical relation $n_c^*(\eta) = n_c^*(0)[\exp(-a\eta^b) + c\eta]$, with 3 free parameters $a$, $b$ and $c$. The fit does not account well for the coexistence and crystalline phase, since the critical particle density displays bumps directly at the ends of the coexistence region. Note that the jammed configuration included lies also on the empirical fit, suggesting that nonequilibrium states are merely an extrapolation of the fluid phase. Interestingly, for the corresponding two-dimensional case the critical particle density displays a minimum and increases again upon approaching the phase transition [5].

## B. SIMULATION OF THE TRACER MOTION

For a single tracer particle exploring the void space, we have computed the mean-square displacements by event-driven molecular dynamics simulations using different host structures (see Fig. 3 of the main text). For each density of the host structure we have generated 32 independent configurations, (16 for the jammed configuration), for each of which 24 trajectories have been calculated with random initial positions on the percolating cluster of the void space. To obtain accurate estimates for the critical exponent $d_w$, the data span 8 non-trivial decades in time, thereby the trajectories cover typical tracer excursions up to, but not exceeding $100\sigma$.

The tracers undergo specular reflections from the obstacles and follow "ballistic" trajectories, i.e., straight lines connecting the collision points. Since energy is conserved, the magnitude $v$ of the velocity remains constant, setting our unit of time $t_o = \sigma/v$. The particle numbers of the hard-sphere liquids were such that the linear size $L$ of the simulation box is always $\approx 200\sigma$ rendering finite-size effects negligible.

In the case of Brownian tracers, the ballistic trajectory the ballistic is interrupted after fixed time intervals $\tau_B$ and a random velocity is assigned according to a Maxwell distribution [6]. On time scales $t \gg \tau_B$, this generates a Brownian motion with diffusion constant $D_o = v^2\tau_B/2d$, also in the presence of one hard-core obstacle. Then, we adopt $t_o = \sigma^2/D_o$ as unit of time.

## C. MEAN-SQUARE DISPLACEMENTS

The local exponents $\gamma_\infty(t) = \mathrm{d}\log\delta r_\infty^2(t)/\mathrm{d}\log t$ directly at the critical density $n_c^*$ shown in Fig. 4 of the main text have been extracted from the data for the mean-square displacements displayed in Fig. S2. The data for the ballistic dynamics display an initial growth $\delta r_\infty^2(t) = v^2 t^2$ until the first collision with an obstacle occurs. Then the mean-square displacements slows down to reach the asymptotic subdiffusive growth $\delta r_\infty^2(t) \sim t^{2/d_w}$ with walk dimension $d_w = 4.81$. The corresponding data for Brownian dynamics initially increase linearly $\delta r_\infty^2(t) = 6 D_o t$, where $D_o$ denotes the short-time diffusion coefficient, the long-time behavior is again subdiffusive, yet with the smaller walk dimension $d_w = 4.24$. The slower initial growth and faster long-time increase of the Brownian dynamics implies that the two mean-square displacements $\delta r_\infty^2(t)$ always cross, see Fig. S2, no matter how the microscopic velocity $v$, respectively the short-time diffusion coefficient $D_o$ is chosen. This crossing is also observed in the local exponent $\gamma_\infty(t)$, see Fig. 4 of the main text.

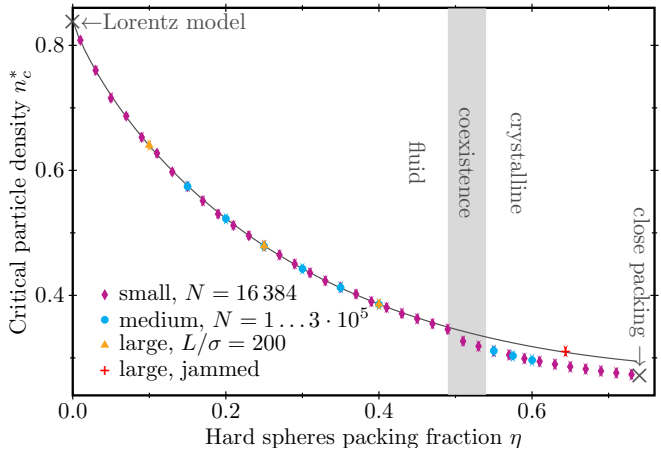

Figure S1. The critical particle density $n_c^* = (N/L^3)\sigma^3$ characterizing the percolation threshold as a function of the hard-sphere packing fraction $\eta$ for different system sizes. The smallest systems correspond to fixed particle number $N = 16\,384$, the medium-sized systems contain $N = 1\ldots 3 \cdot 10^5$ obstacles, while the largest ones, discussed in the main text, correspond to a fixed box size $L \approx 200\sigma$ equivalent to $N = 3\ldots 5 \cdot 10^6$. The solid line is an empirical fit model described in the text and serves as a guide to the eye.

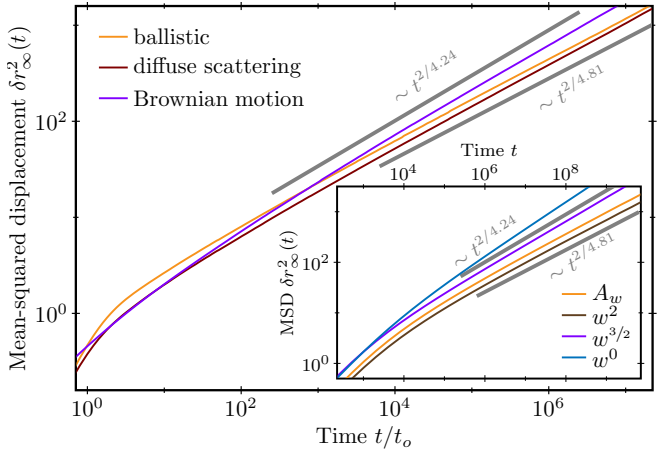

Figure S2. Mean-square displacements $\delta r_\infty^2(t)$ in the Lorentz model ($\eta = 0$) at the critical obstacle density for different tracer dynamics: ballistic motion with specular and diffuse scattering; pseudo-Brownian motion with interrupt times $\tau_B = 0.005 t_o$. Inset: Mean-square displacements $\delta r_\infty^2(t)$ for the hopping motion on the Voronoi network. Here, $t$ denotes the number of steps. Hopping rates are chosen *(i)* proportional to the measured channel cross sections $A_w$, *(ii)* proportional to $w^2$, *(iii)* proportional $w^{3/2}$, and *(iv)* constant as expected for the lattice random resistor network.

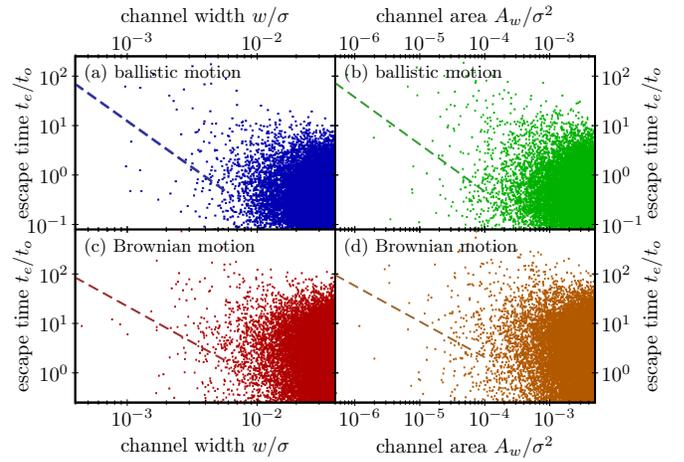

Figure S3. Samples of measured Delaunay cell escape time as a function of (a,c) the width $w$ or (b,d) the cross sectional area $A_w$ of the channel the tracer particle left through. In (a) and (b), the tracer performs ballistic motion, while for (c) and (d) pseudo-Brownian tracer motion was simulated by resampling the tracer velocity after Brownian time steps $\tau_B$. For reference, fits of the typical escape times as obtained in Fig. S5 are included as dashed lines.

The speed-up of the long-time dynamics is at first glance counter-intuitive, yet is fully explained by the scaling behavior of the transition rates. For example in the Lévy–Lorentz gas [7, 8] and in single-file diffusion [9–11], changing the microdynamics from the directed ballistic to the meandering Brownian motion slows down the long-time transport properties. In the Lorentz model, the transition rates vanish faster in the case of ballistic dynamics, $\Gamma \sim w^{d-1}$, than for the Brownian one, $\Gamma \sim w^{d-3/2}$. Thus for large systems there will be always channels so narrow that the ballistic tracer is trapped much longer than the Brownian one. Since the smallest channels dominate the transport properties, the long-time dynamics of the ballistic particles is the slower one.

### D. ESCAPE TIMES

The splitting of the dynamic universality classes originates from the different behavior of the passage times through narrow channels for ballistic and Brownian tracers. As argued in the main text, for ballistic tracers the transition rate $\Gamma$ should scale with the channel cross-sectional area $A_w$ in $d = 3$ as $\Gamma \sim A_w \sim w^2$ as $w \to 0$, where $w$ is the width of the channel. For Brownian tracers, the conductivity of the narrow channels should approach a constant, then the transition rate is anticipated to behave just as an Ohmic conductor to follow $\Gamma \sim A_w / L_w$ where $L_w$ is the length of the channel. Since $L_w \sim \sqrt{w}$ for narrow channels, the Brownian tracers should display transition rates $\Gamma \sim w^{3/2}$ as $w \to 0$.

Here, we examine the first passage times directly to further corroborate the splitting of the dynamic universality classes. For this purpose, we have used a simulation equivalent to the Lorentz model simulation described in the main text: the tracer particle starts from a random point on the percolating cluster of the void space and random initial direction, but have stopped as soon as it leaves the local neighborhood of the 4 adjacent obstacles. This local cell is identified as the Delaunay simplex [12, 13] the particle's start point is located in. We have measured the escape time $t_e$ after which the tracer leaves the Delaunay simplex for the first time using either ballistic or (pseudo-) Brownian tracer dynamics. Simulations have been performed near the percolation transition using 100 random obstacle configurations of linear system size $L = 250$ and dimensionless obstacle density $n_c^* = 0.8375$. In each configuration, $10^6$ initial tracer conditions have been studied, resulting in $10^8$ samples of the escape time in total. Figure S3 shows a subset of these samples with escape time $t_e$ plotted as a function of either the channel width $w$ or the channel cross-sectional area $A_w$ for both the ballistic and Brownian case. The channel width $w$ is defined as the radius of the largest particle which could pass the gap between three spheres of radius $\sigma$ at the corners of the simplex. The cross-sectional area $A_w$ has been obtained by subtracting the area covered by the obstacles from the respective face of the Delaunay simplex.

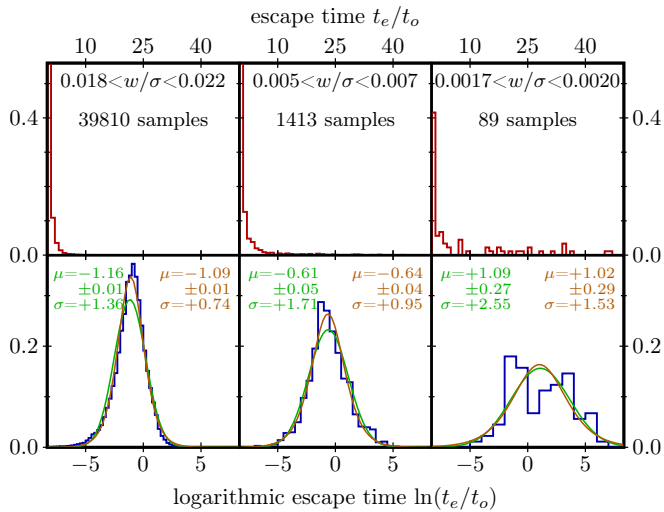

Figure S4. Histograms of single cell escape times $t_e$ for 3 selected channel size bins in the case of ballistic tracer dynamics. Top row: Histograms on linear scales. Bottom row: Same samples using a logarithmic scale in the escape time. The histograms show a broad symmetric peak with increasing width and location for decreasing channel width. Fits to a log-normal distribution are shown in green, orange lines correspond to fits to the log-logistic distribution. Fit parameters are given as text in the respective colors.

A wide range of escape times $t_e$ is observed in Fig. S3 without an obvious trend for small channel sizes $w$. Therefore, we rely on a statistical analysis to obtain a *typical* escape time as a function of $w$. We have binned escape time samples on a logarithmic grid in $w$ with a logarithmic bin width of 0.2, i.e. bin centers are separated by a factor of exp(0.2). Only bins containing at least 12 samples are considered in the following, and the bin width is chosen such that the number of bins included in the analysis is maximized. In order to identify a typical escape time for each bin, we first look at the distribution of individual samples by generating histograms of selected bins. In Fig. S4, histograms of 3 different bins are shown. One infers that the data are well-described by a log-normal or log-logistic distribution of escape times. Similar fits are obtained also for the Brownian case (not shown). As a maximum is clearly observed in the escape time distribution on the logarithmic scale, it is well-founded to use the location of this peak to obtain the typical escape time $t_e$. We have employed the `fitdistr` function of the `MASS` [14] package to fit a normal distribution to the logarithm of the histograms and assign $t_e = \exp(\mu)$ using the mean (and median) $\mu$ of the fitted normal distribution. Let us note that in terms of the log-normal distribution, $\exp(\mu)$ represents the median, while the arithmetic mean would be calculated as $\exp(\mu + \sigma^2/2)$, where $\sigma$ is the standard deviation of the fitted normal distribution. As the standard deviation $\sigma$ also shows scaling with $w$, the mean escape times behave rather differently compared to the typical escape times, which reflects that the distribution displays fat tails. Furthermore, when we consider transition rates $\Gamma \sim t_e^{-1}$, the median of the transition rates corresponds directly to the median of the escape times, in contrast to the arithmetic mean.

Results for the typical escape times $t_e = t_e(w)$ obtained are shown in Fig. S5. Both ballistic and Brownian tracer dynamics is shown and data have been binned either using the channel width $w$ or an effective circular channel width $\sqrt{A_w/\pi}$ using the calculated channel cross-sectional area $A_w$. For channel widths $w$ smaller than one percent of the exclusion radius $\sigma$ the typical escape times grow rapidly as smaller channel width are considered. Furthermore, the growth is significantly faster for ballistic motion than for Brownian motion. For very small channel widths the statistics becomes poor, since there are very few cells in the system with such narrow channels. Yet, in the regime $10^{-3} \lesssim w/\sigma \lesssim 10^{-2}$ a power law can be fitted reasonably well to the data. The obtained exponents are compatible with the theoretical prediction for the *typical* transition rates, $\Gamma \sim t_e^{-1} \sim w^2$ (ballistic) and $\Gamma \sim t_e^{-1} \sim w^{3/2}$ (Brownian), respectively, and the error bars in the exponents are significantly smaller than the difference of the two, fortifying our explanation of the splitting of universality classes of anomalous transport given in the main text of this article.

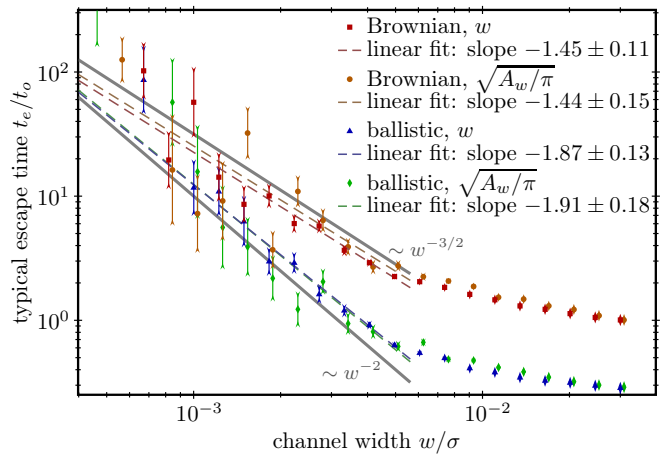

Figure S5. Typical escape time $t_e$ of a tracer particle exiting from a Delaunay simplex through a narrow channel of effective width $w \ll \sigma$ using either ballistic or Brownian dynamics. Included are the typical escape time in terms of an effective circular channel width $\sqrt{A_w/\pi}$ (red and orange data points) obtained from the measured cross-sectional areas $A_w$. Error bars are obtained from the fit uncertainties to the log-normal distributions to the histograms of Fig. S4. The dashed lines represent power law fits $w^b$ to the data points in region of channel sizes $w < 0.0055\sigma$, the exponents $b$ are given in the figure legend. A clear splitting between the scaling of the typical escape time for ballistic and Brownian motion emerges and the fitted exponents differ by 0.5. For comparison, the expected power laws with exponents $-2$ (ballistic) and $-3/2$ (Brownian) are indicated by solid gray lines.